%% file: header_unblinded.tex
\documentclass[acmtog, nonacm]{acmart}
\usepackage{booktabs} 
\usepackage[ruled]{algorithm2e} 

\SetAlFnt{\small}
\SetAlCapFnt{\small}
\SetAlCapNameFnt{\small}
\SetAlCapHSkip{0pt}
\usepackage{comment}
\usepackage{wrapfig}
\usepackage{graphicx}
\usepackage{tabularx}
\usepackage{fancyhdr}

\begin{teaserfigure}
  \centering
  \includegraphics[width=\textwidth]{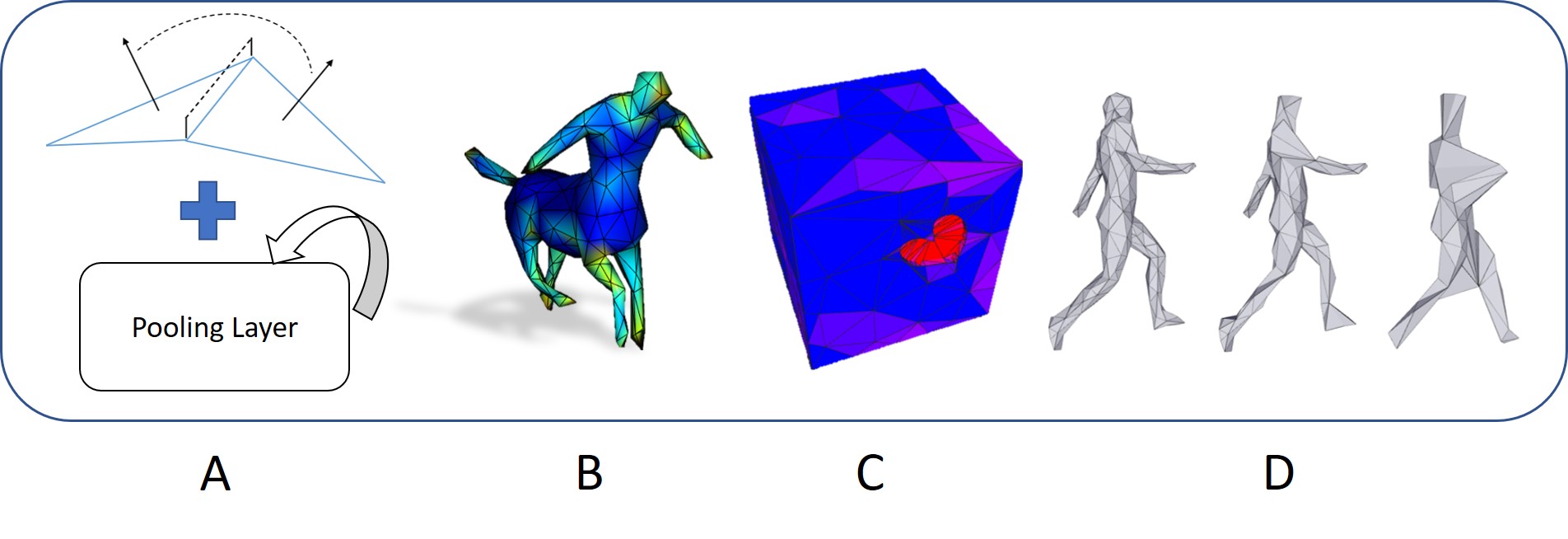}
  \caption{Our Proposed Framework, consisting of a pooling scheme and the Fundamental Forms representation (a), based on the MeshCNN~\shortcite{meshcnn} backbone. Our representation demonstrates high activations in salient and indicative regions (b). Using this scheme, the network indeed focuses on salient regions in deep layers (c). Our Pooling scheme results in more coherent pooling process (d)}
  \label{fig:teaser}
\end{teaserfigure}

\begin{document}
\title{MeshCNN Fundamentals: Geometric Learning through a Reconstructable Representation}

\author{Amir Barda}
\authornote{Both authors contributed equally to the paper}
\affiliation{%
  \institution{Tel-Aviv University}
  \country{Israel}
}
\email{amirbarda@mail.tau.ac.il }

\author{Yotam Erel}
\authornotemark[1]
\orcid{https://orcid.org/0000-0001-8319-5111}
\affiliation{%
  \institution{Tel-Aviv University}
  \country{Israel}
}
\email{erelyotam@gmail.com}

\author{Amit H. Bermano}
\orcid{https://orcid.org/0000-0002-3592-1112}
\affiliation{%
  \institution{Tel-Aviv University}
  \country{Israel}
}
\email{amberman@tauex.tau.ac.il}

\begin{abstract}
Mesh-based learning is one of the popular approaches nowadays to learn shapes. The most established backbone in this field is MeshCNN.
In this paper, we propose infusing MeshCNN with geometric reasoning to achieve higher quality learning. Through careful analysis of the way geometry is represented through-out the network, we submit that this representation should be rigid motion invariant, and should allow reconstructing the original geometry. Accordingly, we introduce the first and second fundamental forms as an edge-centric, rotation and translation invariant, reconstructable representation. In addition, we update the originally proposed pooling scheme to be more geometrically driven. We validate our analysis through experimentation, and present consistent improvement upon the MeshCNN baseline, as well as other more elaborate state-of-the-art architectures. Furthermore, we demonstrate this fundamental forms-based representation opens the door to accessible generative machine learning over meshes.
\end{abstract}
 
\maketitle
\input{main}

\end{document}

%% file: main.tex
\section{Introduction}

What is a good representation for 3D geometry? This question stands at the core of the Computer Graphics field. A meaningful representation leads to powerful solutions to shape-related task, such as retrieval, analysis, manipulation, and generation. As in many other fields, the revolution of Deep Learning has led to many advancements, due to neural networks' semantic understanding and generalization capabilities. Since geometric deep learning is highly non-trivial \cite{7974879}, several recent research directions have emerged, employing different geometric representations \cite{nerf, deepsdf, pointnetplusplus, meshcnn, voxnet, 3dunet}. 

We focus our discussion on learning over 2-manifold triangular meshes. Triangle meshes are the most popular choice for rendering and geometry processing pipelines. Their sparse yet continuous surface representation offers completeness and scalability. While currently not displaying state-of-the-art performance for shape analysis tasks, this representation has drawn notable attention over the last few years \cite{meshcnn, meshwalker, hodgenet, pd_mesh_cnn, dualprimalGCN}. Perhaps the most seminal work in the field is MeshCNN~\cite{meshcnn}. This work defines edge-centric convolution and pooling operations for meshes, along with a set of features that describe the mesh for each edge. This work has inspired many follow-up research, proposing anything from novel attention layers \cite{pd_mesh_cnn}, to employing the network's concepts for mesh editing tasks \cite{neural_subdiv, texture_synth}.

In this paper, we analyze the way geometry is represented throughout the MeshCNN network, and introduce higher quality learning through geometric reasoning. 

We start by experimenting with the five features proposed by the MeshCNN framework that describe the geometry, and quickly realize that, as can be expected, not all features are created equal. 
Through these experiments, along with investigations of explicit and differential coordinates, we submit that features fed to the network should have two key ingredients: they should be rotation and translation invariant, and should be reconstructable, i.e., should provide enough geometric description that at least allows for the reconstruction of the original geometry. As demonstration, we propose the minimal yet elegantly fitting representation of the first and second fundamental forms \cite{Crane2005DiscreteDiffGeo} --- namely the length and dihedral angle of each edge.  As we demonstrate in Section~\ref{sec:results}, these features yield higher accuracy for most of the popular mesh analysis tasks. In addition, since the fundamental forms naturally lend themselves better to geometric functions, they enable learning reconstructable representations, or in other words, they open the door to better mesh generation capabilities. We demonstrate this claim through de-noising experiments, where the network is asked to reconstruct a noisy input mesh through a vanilla auto-encoder paradigm.

We argue that since the network should understand 3D geometric shapes, equipping it with geometry-driven processes acts as powerful guidance. Under this light, we inspect the pooling operation as well. Observing that the pooling layer's decision making process does not follow geometric reason, we adapt its scheme to achieve consistent performance improvement across all our experiments.


Our main contributions can be summarized as follows:
\begin{itemize}
\item Introduction of the first and second fundamental forms as a representation fitting for learning meshes.
\item Analysis of the properties this representation holds that lend themselves to learning shape analysis tasks, as well as generative ones.
\item Enhancing the pooling layer to update the mesh after each edge collapse, thus better preserving the accumulated information throughout the process.

\end{itemize}

\section{Related Work}

The success of deep neural networks in the realm of 2D image processing and machine vision has inspired many to investigate applying these techniques to various tasks over 3D geometry. The differences between the various approaches primarily stem from the geometric representations used by each one. Next we briefly introduce the works done along these axes. For an in depth survey of recent work in geometric deep learning, we refer the reader to \citet{gezawa2020review}.

\paragraph{Image-Based Representations.} Perhaps the most natural way to apply established 2D methods to 3D data is to render the geometry into a set of 2D images, and learn from them. The first works rendered the geometry from several directions and stacked the results together for a network to learn \cite{boulch2017unstructured,feng2018gvcnn,kanezaki2018rotationnet,yavartanoo2018spnet,qi2016volumetric,bai2016gift,he2018triplet}. Later approaches added other modalities, such as depth maps or planar parameterization techniques in order to add more geometric meaning to the 2D images \cite{GWCNN,gomez2017lonchanet,sarkar2018learning,zanuttigh2017deep,haim2019surface,maron2017convolutional}. While still an active field of research \cite{han20193d2seqviews,wang2019dominant}, we argue that 2D images are inherently unfit to represent 3D geometry. Their regular sampling rate does not represent different frequencies well, e.g. gaps or sharp corners. In addition, proper representation of a geometry typically requires several images, which burdens the process in terms of computational load and memory footprint. 

\paragraph{Volumetric Representation.}
Another natural approach is to extend the established 2D convolutions neural networks employ so well to 3D ones. This typically means equally dividing the space into a 3D grid, or a \textit{voxel} grid, and employing 3D convolutions over the regularly sample data \cite{wu20153d,sedaghat2016orientation,brock2016generative,wang2019normalnet,deepsdf,3dunet}. While this approach is a better fit for 3D geometric, as it is inherently structured to reflect 3D relations, it still suffers from the curse of dimensionality. An image of 4K resolution (i.e., roughly eight million pixels) bears the same amount of elements as a $200 \times 200 \times 200$ voxel grid. Hence, the complexity of voxel grids inhibit their practicality. Some works have started exploiting tree structures to significantly reduce memory footprint through adaptive cell sizes \cite{roynard2018classification,10.1145/3072959.3073608,10.1145/3272127.3275050}. These works essentially seek to balance the networks' need for regular sampling with the data's typical irregularity. However, as of now no such method has presented a practical, scalable and highly accurate model the is successful in finding this balance.

\paragraph{Point clouds.} Point clouds are the currently leading representation in terms of accuracy of the models that employ them. Points clouds are sets of points (and optionally normals) that reside on the objects boundary. Their dense sampling rate lends itself naturally to the neural networks' preference for regular data, while the fact that they only reside on the surface provides sparsity. The challenge pointclouds hold for neural networks is that they are unordered, implying that the learned model should be permutation invariant. PointNet \cite{qi2017pointnet} was the first to introduce a permutation invariant model for geometric learning of pointclouds. Since then this core concept has been utilized and developed further many times, employed for tasks such as classification and segmentation \cite{pointnetplusplus,ben20183dmfv,atzmon2018point,li2018pointcnn,thomas2019kpconv,tchapmi2017segcloud,guerrero2018pcpnet,liu2019relation,wang2019graph,engelmann20203d,Liu_2019_ICCV}, but also for generation \cite{wang2019graph,Shu_2019_ICCV,Li_2019_ICCV}. Even though the state-of-the-art in shape analysis and generation lies with this representation, we argue that it still holds inherent drawbacks. First, even  though its dense sampling rate is orders of magnitude more efficient than voxel grids, it is still wasteful. More often than not, shapes are composed of regions with much variation and features, but also with large regions that are flat or are otherwise less informative. A geometry-aware adaptive sampling rate is something current networks are not able to handle. Secondly, pointclouds do not describe connectivity. Typically in pointclouds, connectivity is implicitly assume through local proximity to other points. While this prior is usually accurate, it fails to describe well highly curved regions, or different parts of the shape that are very close to each other. Meshes, on the other hand, offer solutions to these two drawbacks.

\paragraph{Graph Representations.} A natural approach to learning on 3D triangular meshes is to view them as graphs, which allows employing variants of GCNs (Graph Convolutional Networks) \cite{henaff2015deep, kipf2017semisupervised, simonovsky2017dynamic, dualprimalGCN,mo2019structurenet}. These works define kernels that operate typically over each vertex and its 1-ring neighborhood, and compensate their weights according to vertex valence and edge lengths. However, graphs alone consider only the vertices and edges of the mesh for the learning process. They do not consider perhaps the most important aspect of the mesh representation, which are the faces. Hence, graph-based methods that better accommodate for the connectivity information that the mesh has to offer typically show preferable performance to purely graph-based ones.

\paragraph{Mesh Representations.}
Convolutions that exploit mesh connectivity structures were first introduced by \citet{masci2015geodesic}. Since then, several approaches addressed the irregularity of the mesh structure and sampling rate, proposing ideas such as sampling the neighborhood of each vertex uniformly, or achieving regularity through spectral decomposition \cite{boscaini2016learning,poulenard2018multi,gong2019spiralnet++,lim2018simple,schult2020dualconvmesh, zhou2020fully, verma2018feastnet}.
The seminal work of MeshCNN~\cite{meshcnn}, which we base our investigation on, proposed a novel convolution and pooling operations that exploited the mesh structure, yielding a rotation invariant, simple, and highly accurate shape analysis architecture. Using the ideas of this work as a backbone, several publications have applied this engine to downstream applications, such as mesh super-resolution and geometric texture synthesis \cite{hanocka2020point2mesh,neural_subdiv}. 

Being innovative, MeshCNN is a proof-of-concept, and is not thoroughly optimized in terms of architecture and features. For example, follow-up work has suggested adding attention layers to it ~\cite{pd_mesh_cnn}. While we compare our accuracy to the latter work, we argue it is orthogonal to the claim of this paper, as one could also add additional layers and teaks to the system we propose. Furthermore, unlike GCNs, MeshCNN and its followup cannot currently perform tasks which require recovering the mesh geometry at the output - such as de-noising. Another recently proposed novel approach suggests accumulating geometric features over a random walk performed over a mesh using an RNN paradigm \cite{meshwalker}. This approach poses several advantages, such as low memory footprint, accuracy and robustness. However, besides the burden of training the notoriously hard to train RNN architecture, this approach can not be used for generative downstream tasks, and because the machinery described works for any type of graph, it is less geometrically motivated and fails to exploit strong priors existing in triangle meshes like regularity of the edge or face neighborhoods. Hence again, we compare our performance to this work, but again make the same claim of orthogonality. 
Lastly, we note the difference between this line of work, which we belong to, that seeks to understand the structure and connectivity of the mesh, to the approaches that deform the same mesh throughout the process (i.e., networks that learn to manipulate the geometry of a constant connectivity) \cite{yuan2020mesh,zhou2020fully,gupta2020neural}. Most real work application cannot be efficiently described using a constant connectivity.

\section{Preliminaries}

\subsection{MeshCNN Review}
\label{sec:meshcnn_prelim}

\begin{figure}
    \centering
    \includegraphics[width=.99\columnwidth]{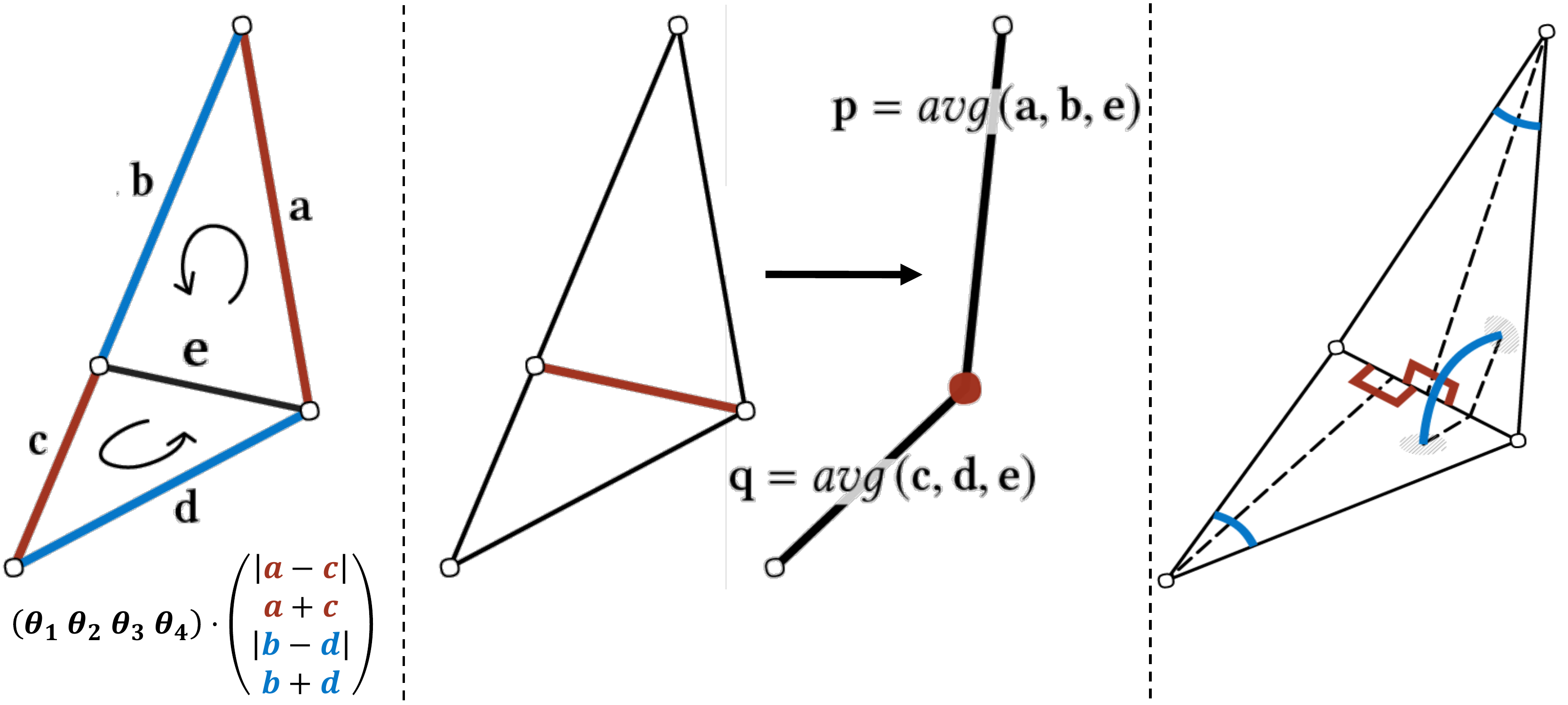}
    \caption{MeshCNN definitions (figures from \citet{meshcnn}). Left: a convolution centered at edge $e$ is defined as formulated below. Note the operator is symmetric with regards to the blue and the red edges. Middle: edge pooling operation, the five edges of the 1-ring become two, with averaged feature values. Right: the input representation. Each edges is described through 5 numbers. The dihedral angle (blue), the two opposite angles (blue), and the two edge length-to-triangle height ratios (red). }
    \label{fig:meshcnn_rulez}
\end{figure}
For the sake of completeness, we give a brief overview of the MeshCNN pooling and convolution layers\cite{meshcnn}. For full details, we refer the reader to the original paper.

\subsubsection{Convolutions}
MeshCNN is based on the observation that every edge in a watertight 2-manifold mesh has a fixed amount of incident faces. Therefore, unlike previous art, MeshCNN defines the convolution on the edges of the mesh. For every edge $e$, the convolution is performed on its 1-ring neighborhood (i.e., on all edges of the incident triangles), in counter clockwise order (see Figure~\ref{fig:meshcnn_rulez}, left). This creates an ambiguity as the order can be either $(a,b,c,d)$ or $(c,d,a,b)$. To resolve this, The convolution operator is defined to be the order invariant functions: 
\begin{equation}
  \begin{aligned}
    & e\cdot \theta_0 + \sum_{j=1}^4{\theta_j \cdot e^j}, \\
    & (e^1,e^2,e^3,e^4) = (|a-c|,a+c,|b-d|,b+d),
  \end{aligned}
\label{eq:convolution}
\end{equation}
where $\theta_i$ are the learned weights.

\subsubsection{Pooling}
One of the distinguishing features of MeshCNN is its pooling layer, which collapses edges according to their feature norms. A strong feature norm signals an area with important information for the next layer, so the pooling layer chooses the weakest ones and removes them according to a simple mesh simplification scheme~\cite{10.1145/258734.258843}. After all the edges to-be-pooled are chosen (i.e. the target number of edges is reached), the pooling layer updates to features on the edges by averaging the features in the collapsed 1-ring, as shown in Figure~\ref{fig:meshcnn_rulez}, middle. 

\subsubsection{Representation}
As portrayed in Figure~\ref{fig:meshcnn_rulez}, right, each edge is represented by five features: 
\begin{itemize}
    \item The angle between the normal vectors of its two incident faces.
    \item The two angles opposite to the edge.
    \item The two ratios between the edge length and the corresponding triangle's height.
\end{itemize}
These features are invariant to rotation, translation and uniform scaling (although scaling invariance can always be achieved through bounding box normalization). 

\subsection{Fundamental Forms}
Intuitively, the first fundamental form $I$ measures lengths on the surface induced by the euclidean metric, and the second fundamental form $II$ measures the rate of change of the normal field. The two fundamental forms are established measures, known to be coordinate invariant and to fully characterize the shape up to translation and rotation.
In the continuous case, for a surface $M$ with a parameterization $(u^1, u^2)$, if there is a local embedding $x$ in $R^3$, $I$ can be induced by $I_{\alpha \beta} = (x_{,\alpha},x_{,\beta})$ and $II$ is induced by $II_{\alpha,\beta} = (x_{,\alpha \beta},N)$, where $x$ is a position on the surface, $\alpha, \beta = {1,2}$ are indices of the parameterization, and an index after comma denotes a partial derivative with the respect to the index. 
In the discrete case of a triangular mesh, it turns out that the fundamental forms conveniently reside on mesh edges, are simply the edge length, and the edge dihedral angles respectively \cite{Wang2012LinearSurfaceRecon}. 

\section{Method}

The premise of this report is that since the network is expected to learn geometry, imposing upon it geometrically consistent operations would offer powerful guidance for the training process. We investigate and apply this principle to two points in the MeshCNN framework --- its pooling operation and its geometric representation. In the following, we describe the investigation, the conclusion, and its application for each of the two aspects. 

\subsection{Representing Geometry}
\label{sec:representation}

The investigation begins with examining the features proposed by the original work (see Section~\ref{sec:meshcnn_prelim}). As a test-bed we inspect the task of mesh classification, using the SHREC dataset \cite{shrec}. The performance of various combinations of the proposed features are listed in Table~\ref{tab:meshcnn_features_combi}. The results clearly indicate that the features are not equal. The opposite angle features yield the best accuracy, while the other two yield lower scores. In contrast, we that the dihedral angle is more effective in conjunction with the other two features, while they do not contribute to one another. Following the main premise of geometric reasoning, we observe that the effects on accuracy corresponds to geometry reconstructability, i.e, features that allow recovering the mesh geometry will perform better. For this reason, the opposite angles and length ratios do not contribute to each other; since they accordingly do not contribute much to reconstructing the geometry. We hence postulate that reconstructability is a good measure of the potential information the features contribute to the learning process. We argue that features that provide full \textbf{reconstructability} also provide useful information to the network. Indeed, inspecting the behavior of the originally proposed features reinforces the suspicion that these non-reconstructable features convey weaker signals to the learner. Attacking this from another angle, we visually inspect the spatial distribution of the information conveyed by the originally proposed features. Figure~\ref{fig:proposed_features} top, portrays the $L_2$ norms of the five features on each edge. No observable pattern can be seen in regions that are expected to be unique or indicative. This again cues that these features are less informative for the network, in correlation to reconstructability. 
\begin{table}%
\caption{Feature Combinations. The accuracy results on SHREC test set using various features.}
\label{tab:meshcnn_features_combi}
\begin{minipage}{\columnwidth}
\begin{center}
\begin{tabular}{ll}
  \toprule
  Method   &  Accuracy \\  \midrule
  Dihedral angle     & $85.0\%$ \\
  Length ratios    & $85.0\%$ \\
  Opposite angles   & $90.83\%$ \\
  Opposite + ratios   & $91.67\%$ \\
  Dihedral + ratios   & $96.6\%$ \\
  Dihedral + Opposite  & $95.0\%$\\
  MeshCNN   & $\mathbf{98.6\%}$ \\
  \bottomrule
\end{tabular}
\end{center}
\bigskip\centering
\end{minipage}
\end{table}%

\begin{wrapfigure}[8]{h}{0.25\textwidth}
    \vspace{-0.5cm}
    \includegraphics[width=0.15\textwidth]{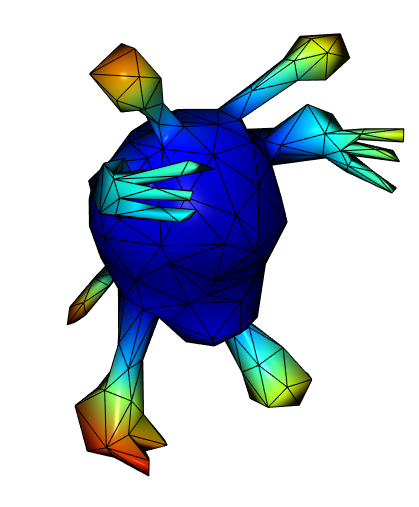}
  \label{fig: selfRivetingExample}
\end{wrapfigure}

We therefore propose using more traditional reconstructable features, such as the coordinates themselves, or the widely established Laplacian coordinates. Laplacian coordinates are vectors which give the difference between the vertex and its neighbors' (possibly weighted) average, and are employed in countless geometry processing operations \cite{egst.20051044}. By definition, the norm of these coordinates is directly linked to the local rate of change (See heat-map in inset). The results of these experiments over the same aforementioned setup are depicted in Table~\ref{tab:xyz_laplace}. In contradiction to the aforementioned suspicion, these reconstructable features do not yield better performance. Our insight is that the reason is rotation and (or) translation invariance. These transformations are natural symmetries for the task, since a rotated chair is still as it perceived by us. Indeed, as can be seen, when adding random rotations to the test and training process, the performance of these features is lowered. Even more so, we see that when converting the explicit coordinates to rotation invariant ones, simply by taking their dot products and average of norms instead of a simple average, accuracy is dramatically improved. This is even though the latter operations reduce how descriptive the representation is. This strongly indicates that \textbf{rotation and translation invariance} is a necessity. 

To validate this claim, we propose the slimmest representation that is both rotation and translation invariant, and is reconstructable --- the first and second fundamental forms. Indeed, when examining this representation's spatial behavior, we can observe a strong correlation between it and salient and distinctive regions in the mesh (see Figure~\ref{fig:proposed_features} bottom). As we demonstrate in Section~\ref{sec:results}, these two numbers alone induce higher quality learning than the alternatives. In addition, we also demonstrate that since this representation is reconstructable, it naturally lends itself to tasks of reconstruction or generation. 

We note that adding more features or architectural elements to framework would probably improve the learning even further, however this investigation is out of the scope of this paper, and is left for future work.

\begin{table}%
\caption{Performance of various reconstruction-abling features on SHREC.}
\label{tab:xyz_laplace}
\begin{minipage}{\columnwidth}
\begin{center}
\begin{tabular}{ll}
  \toprule
  Feature Type   &  Accuracy \\  \midrule
  xyz coordinates    & $76.73\%$ \\
  xyz w/ random rotations   & $69.78\%$ \\
  xyz dot and norms    & $82.85\%$ \\
  Laplacian coordinates & $54.44\%$\\
  MeshCNN   & $\mathbf{98.6\%}$ \\
  \bottomrule
\end{tabular}
\end{center}
\bigskip\centering
\end{minipage}
\end{table}%

\subsection{Pooling}

\begin{figure}
  \includegraphics[width=0.99\columnwidth]{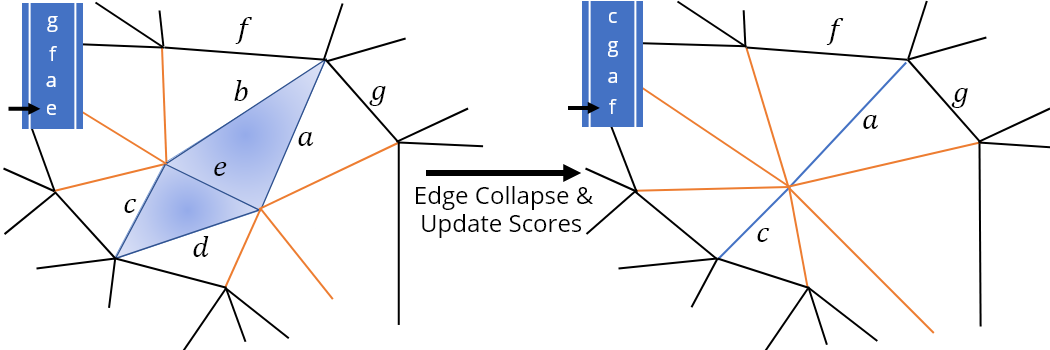}
  \caption{The enhanced pooling layer. Left: an edge $e$ with its 1-ring neighborhood $\{a, b, c, d\}$ was selected to be collapsed due to its low features norm. Notice that edge $a$ had the second lowest score prior to collapse. Orange edges were marked to emphasize the fact they will also be affected by the collapse geometrically, but we do not take this into account when computing the new scores. Right: After collapsing $e$, we calculate the new scores for $a$ and $c$ (by averaging upon the old $a, b$ and $e$ for $a$, and the old $c, d$ and $e$ for $c$) suddenly $a$ doesn't have the lowest score, and instead some other edge $f$ will be selected next. This is on contrary to the original layer which would have selected $a$.}
  \label{fig:pool_patch}
\end{figure}

\begin{figure}
  \includegraphics[width=.99\columnwidth]{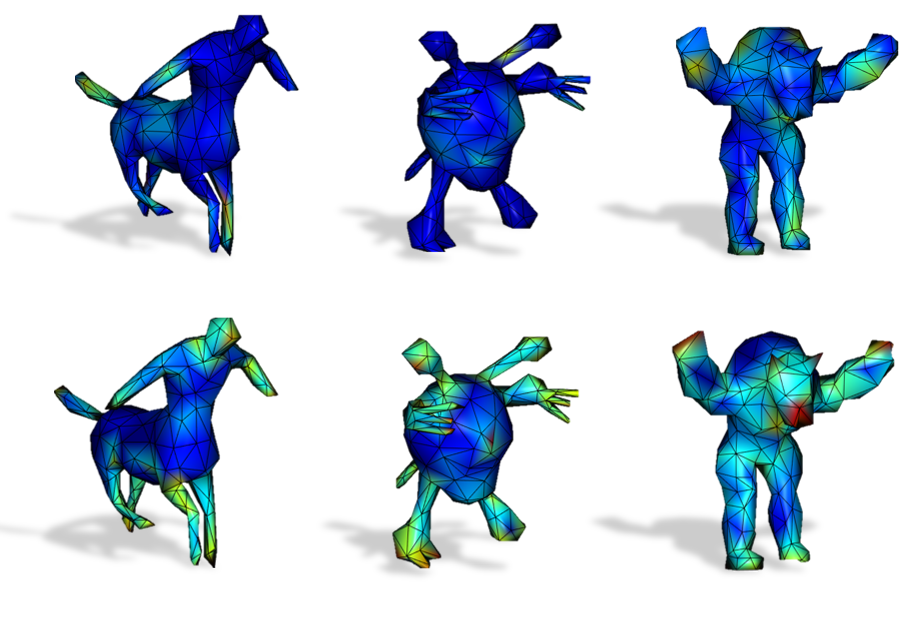}
  \caption{Feature norm comparison between the original 5 edge features in \cite{meshcnn} (top) and our 2 proposed edge features (fundamental forms, bottom) on several models from the SHREC16 dataset. All feature channels are normalized by subtracting the mean and dividing by the standard deviation. MeshCNN's pooling is performed according the feature norms, with edges with lower norms pooled first. The heat maps indicate that our features provide better indication of "areas of interest" such as hands, legs and heads in the shown meshes. This provides a better starting point for the network and improves performance overall, as we observed empirically.}
  \label{fig:proposed_features}
\end{figure}

\begin{figure*}
  \includegraphics[width=\textwidth, keepaspectratio]{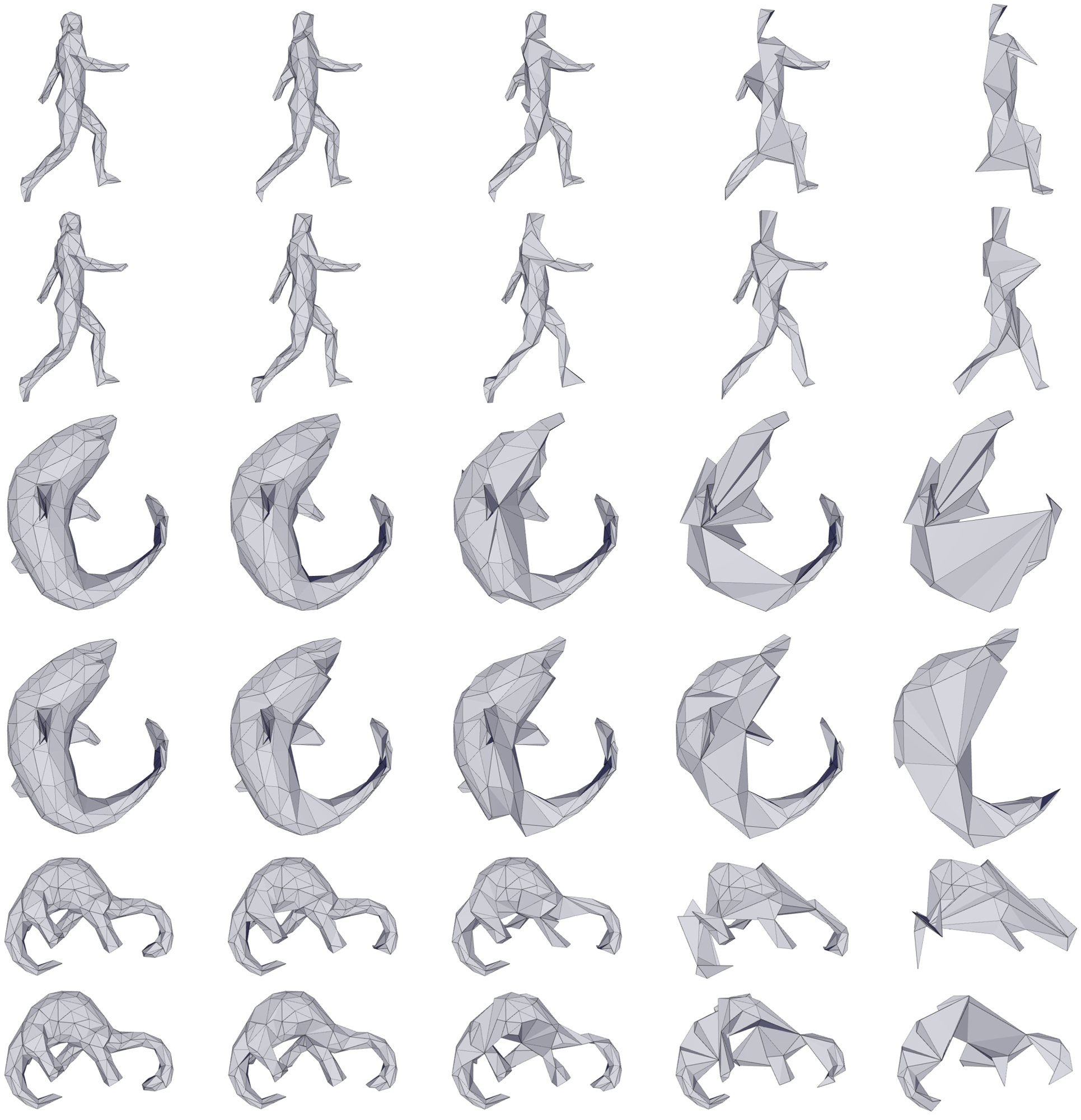}
  \caption{Qualitative evaluation of pooling the edges. From left to right: an original mesh is being pooled into lower spatial dimensions. Top rows: MeshCNN pooling, Bottom rows: our pooling.  By using the trained networks and passing test meshes through them, we can visualize how the pooling layers perform by applying the new topology after each layer to the geometry of the original mesh. By updating the scores of edges during the pooling operation, the global shape is preserved better.}
  \label{fig:pool}
\end{figure*}

The originally proposed MeshCNN pooling layer chooses the $N$ edges with the lowest norm values to collapse, where $N$ is a predefined hyper-parameter. This imposes the popular assumption that informative regions should bear stronger activations. This also reinforces the feature selection process described in Section~\ref{sec:representation} and Figure~\ref{fig:proposed_features}.  The problem with this approach, however, is in the selection mechanism, which does not follow geometric reasoning. As opposed to traditional 2D pooling, where pixels are simply omitted. The pooling operation is designed maintain geometric integrity, by sifting information from the collapsing edge into its neighborhood. However, choosing the $N$ edges to collapse probably implies collapsing many edges in the same area. This contradicts the geometric meaning of the pooling operation. Presumably, after enough edges of a region are collapsed, the aggregated information reaches a certain threshold and should not be collapsed again (see Figure~\ref{fig:pool_patch}).
Hence we propose to update the features after each edge collapse. Since the collapse operation is local, so is the required update after each iteration, making the computational load of the update operation to be negligible. As we demonstrate in Section~\ref{sec:results}, this lightweight step consistently improves accuracy across all our experiments. In addition, when qualitatively following the collapsed edges throughout the network, a clearer representation of the shape can be observed using the updated pooling scheme compared to the original one, as depicted in Figure~\ref{fig:pool}. This indicates the model is more attentive to the entire shape during classification, suggesting robustness to disturbances.

\section{Results}
\label{sec:results}
We perform several experiments to validate our results. We employ our system for classification on the SHREC \cite{shrec} and Cube engraving\cite{meshcnn} datasets. We then perform segmentation on the Human Segmentation dataset prepared in an identical fashion as \citet{meshcnn}. Importantly we use identical network architectures that were used by \citet{meshcnn}, to isolate the effects of the discussed aspects alone (i.e., the representation and pooling scheme). Additionally, we compare our results to state-of-the-art works \cite{meshwalker, hodgenet, pd_mesh_cnn} when applicable. 
Finally, we perform ablation studies, and demonstrate the ability to perform a generative task of mesh de-noising.

\subsection{SHREC classification}
We first test our performance on the SHREC dataset \cite{shrec}, and similarly to \cite{meshcnn} split the dataset into 16/4 training/test examples per class (Split 16) and 10/10 split as well (Split 10). The total amount of samples is 600. We train using the exact setup and hyper parameters proposed by \citet{meshcnn}, only replacing the features and the pooling layer with ours for fairness. See Table~\ref{tab:results} for summary of results. It is worthy to note that the SHREC classification dataset is relatively well aligned globally and ablation studies performed by \cite{meshcnn} have shown that using naive Cartesian coordinates as features yield quite good results. This would not be the case were the input meshes were randomly rotated. See section \ref{sec:ablation} for more information.
\begin{table}%
\caption{Test set accuracy for various methods and datasets. Human segmentation results are soft edge accuracies.}
\label{tab:results}
\begin{minipage}{\columnwidth}
\begin{center}
\begin{tabular}{lcccc}
  \toprule
  Method   &  SHREC  & SHREC & Cubes & Human \\ 
           &  Split 16 & Split 10 & Engraving & Segmentation\\  \midrule
  Ours     & $\mathbf{100\%}$ & $\mathbf{99.33\%}$ & $\mathbf{99.24\%}$ & $93.23\%$ \\
  MeshCNN  & $98.6\%$  & $91.0\%$ & $92.16\%$ & $92.3\%$ \\
  MeshWalker  & $98.6\%$  & $97.1\%$ & $98.6\%$ & $\mathbf{94.8\%}$\\
  HodgeNet    & $99.17\%$  & $94.67\%$ & - & - \\
  PDMesh   & $99.7\%$  & $99.1\%$ & $94.39\%$ & $91.11\%$\\
  \bottomrule
\end{tabular}
\end{center}
\bigskip\centering
\end{minipage}
\end{table}%

\subsection{Cubes Engraving classification}
This dataset contains 23 classes of shapes embossed into a random face of a cube, consisting of 4600 meshes. Although being somewhat contrived, the connectivity element of the 3d structure plays a much more important role for these sort of meshes (the engraved areas). Attempts to solve this task using a topology agnostic solution yield inferior results. It is clear from the results (Table~\ref{tab:results}) that using our features and pooling layer improves accuracy with respect to \cite{meshcnn}, to the extent that it surpasses all other state-of-the-art as well.

\begin{figure*}
  \includegraphics[width=.99\textwidth]{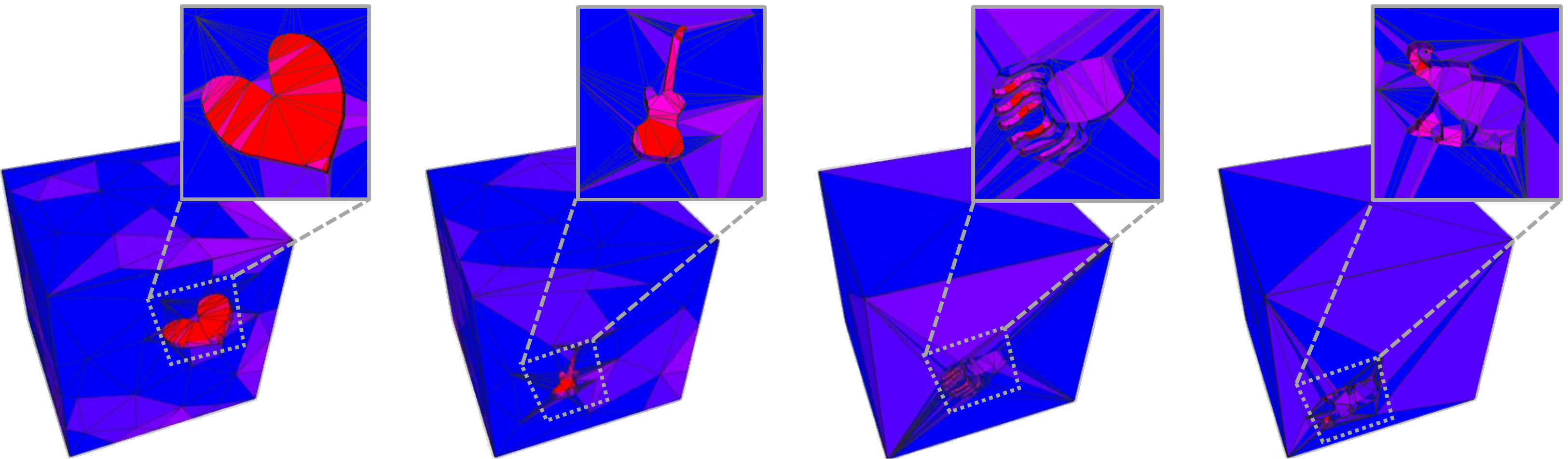}
  \caption{Visualization of our network's shape understanding. During a forward pass on a test mesh from the cubes dataset, we measure when in the process was every edge collapsed. Colder (more blue) edges were collapsed early in the pipeline, while warmer (more red) regions were collapsed later, or not at all. As can be seen, the network learns to focus on distinctive features in the engravings (and not on the engraved cube itself) such as the octopus's legs and the elephant's trunk.}
  \label{fig:shape_understanding}
\end{figure*}

\subsection{Human Segmentation}
This dataset contains 370 models of humans for training from SCAPE \cite{SCAPE}, FAUST \cite{FAUST}, MIT \cite{mit} and Adobe Fuse \cite{adobe}, and the test set consists of 18 models from SHREC07 \cite{shrec07} humans dataset. The models are annotated with a per-face labels segmented using \citet{mesh_seg}. The 3D meshes include humans in various poses, with the same topological properties: 752 vertices, 2252 edges and 1500 faces. Note The meshes are all manifolds with genus zero, with no self intersecting faces. See Table~\ref{tab:results} for results. Note we use soft edge labels as in \citet{meshcnn} that take into account edge lengths. The results for \citet{pd_mesh_cnn} were taken from their supplementary material.

\subsection{Mesh De-noising}
For mesh de-noising, we use the SHREC dataset. In addition, we employ the segmentation architecture of MeshCNN, only instead of outputting $L$ channels that correspond to the $L$ labels of the segmentation, we output the geometry for every edge. We compare using explicit $XYZ$ coordinates as output to our reconstructable fundamental forms. Quantitative results for the test set are reported in Table~\ref{tab:denoising}. As can be seen, learning the fundamental forms is significantly simpler for the network, compared to trying to figure out the explicit coordinates. In fact, the explicit $XYZ$ coordinates even fail to achieve the precision a simple identity reconstruction would have. This results continues reinforcing our rotation and translation invariance claim. In contrast, generating the Fundamental Form representation is significantly simpler for the training process to learn, yielding an error that is 10 times smaller than the identity. The proposed fundamental forms also bear merit over employing the representation proposed by MeshCNN. 

\begin{table}%
\caption{De-noising Results. We applied random Gaussian noise to all vertices with variance 0.1. FF stands for Fundamental Forms, MeshCNN features are described in \ref{sec:meshcnn_prelim} and XYZ are cartesian coordinates. Notice Identity entries indicate what would be the Average MSE between a noisy mesh and its clean counter part (serves as baseline).}
\label{tab:denoising}
\begin{minipage}{\columnwidth}
\begin{center}
\begin{tabular}{ll}
  \toprule
  Experiment   &  Average MSE \\  \midrule
  FF Identity & $0.05$ \\
  FF -> FF     & $0.0096$ \\
  MeshCNN -> FF   & $0.012$ \\
  XYZ Identity & $0.01$ \\
  FF -> XYZ    & $0.08$ \\
  MeshCNN -> XYZ  & $0.082$\\
  \bottomrule
\end{tabular}
\end{center}
\bigskip\centering
\end{minipage}
\end{table}%

\subsection{Feature Validation and Comparisons}

Of course, many choices exist for reconstructable features. In this work we choose the slimmest possible ones, to demonstrate their effectiveness. 
We compare this representation to a few popular alternatives. 

\paragraph {Cartesian Coordinates. }
The most naive choice is to simply use the Cartesian midpoint of each edge as its features, namely, 3 input channels corresponding the x,y and z coordinate value respectively. This was attempted in the original meshCNN paper, and achieved a surprisingly good result of 91\% test accuracy. Upon further examination, we have found this result is largely due to the alignment between meshes in the SHREC dataset. Randomly rotating the meshes during training lowers the training accuracy to 70\%. 

We have also experimented with a rotation invariant version of this representation - through taking the dot product of the two incident vertices, and the average of their norms. Significant improvement can been seen by this seemingly less informative representation, which strongly indicates the need for rotation invariance.

\paragraph {Laplace Coordinates. }
Laplace coordinates $\delta$, are the result of multiplying the Laplacian matrix $L$, with the Cartesian coordinates $V$. Laplacian coordinates are vectors which give the difference between the vertex and its neighbors' (possibly weighted) average.  Nevertheless the coordinates themselves are not rotation and translation invariant, which hinders performance on datasets without mesh alignment.

\paragraph{MeshCNN Features. }
Additionally, we try using several combinations the originally proposed features. The conclusions from these experiments are depicted in Section~\ref{sec:representation}.

Test accuracy results on the SHREC16 dataset for each of these experiments is given in Tables~\ref{tab:xyz_laplace} and \ref{fig:proposed_features}. 

\subsection{Ablation Study}
\label{sec:ablation}
We perform an ablation study that verifies incremental improvement from \citet{meshcnn} by using our enhanced pooling layer (PL) and by using the fundamental forms as features (FF). See table \ref{tab:ablation} for detailed results.

\begin{table}%
\caption{Ablation Study}
\label{tab:ablation}
\begin{minipage}{\columnwidth}
\begin{center}
\begin{tabular}{lcccc}
  \toprule
  Method   &  SHREC  & SHREC & Cubes & Human \\ 
           &  Split 16 & Split 10 & Engraving & Seg.\\  \midrule
  MeshCNN& $98.6\%$ & $91.0\%$ & $92.16\%$ & $92.3\%$\\
  PL & $\mathbf{100\%}$ & ${96.33\%}$ & $96.5\%$ & $91.5\%$\\
  FF & $\mathbf{100\%}$ & $\mathbf{99.33\%}$ & $94.08\%$ & $92.49\%$\\
  Ours (PL, FF)& $\mathbf{100\%}$ & $\mathbf{99.33\%}$ & $\mathbf{99.24\%}$ & $\mathbf{93.23\%}$\\
  \bottomrule
\end{tabular}
\end{center}
\bigskip\centering
\end{minipage}
\end{table}%

\section{Discussion and Future Work}
We presented a representation and geometric reasoning to introduce better learning to the decorated MeshCNN  framework \cite{meshcnn}. Our contributions are in the form of a reconstructable representation and a better pooling mechanism. Another benefit of our work is enabling generative tasks to be carried out using the same backbone, exposing the field of geometric learning over meshes to a new frontier. While MeshCNN certainly performs better under the proposed paradigm, the pooling layer is relatively computationally demanding, and does not scale well to larger models. This is due to the sequential nature of selecting edges from the data structure sorted by the scores, whereas updating the scores of the neighbors has proven to have negligible difference in computational time. We do believe future work must focus on the optimization of the pooling layer, as models in larger scales are almost always represented by a mesh due to their inherent sparsity, which isn't fully exploited by our technique. Another important factor that distinguishes the task of shape understanding using our technique, is the enhanced explain-ability of the decision making made by the network enabled by geometrically motivated operations. For example, tracing the pooled edges back to their parent (Figure~\ref{fig:shape_understanding}) to visualize the networks attention, or being able to add a reconstruction loss term to the training scheme since reconstruction is possible at the output. We believe that the convolution operation itself could also gain from better geometric reasoning, and leave this investigation for future work. Lastly, we argue that the reconstructable features could open up the door to a plethora of generative tasks that could be carried out by the MeshCNN backbone, and are anxious to see what this line of work could yield in the near future.

\bibliographystyle{ACM-Reference-Format}
\bibliography{bibliography}